\documentclass[sigconf]{acmart}

\usepackage{booktabs, makecell, array, balance}

\usepackage{subcaption}

\AtBeginDocument{%
  }

\copyrightyear{2025}
\acmYear{2025}
\setcopyright{cc}
\setcctype{by}
\acmConference[UIST '25]{The 38th Annual ACM Symposium on User Interface
Software and Technology}{September 28-October 1, 2025}{Busan, Republic of
Korea}
\acmBooktitle{The 38th Annual ACM Symposium on User Interface Software and
Technology (UIST '25), September 28-October 1, 2025, Busan, Republic of
Korea}
\acmDOI{10.1145/3746059.3747757}
\acmISBN{979-8-4007-2037-6/2025/09}

\begin{document}

\title{HandOver: Enabling Precise Selection \& Manipulation of 3D Objects with Mouse and Hand Tracking}

\author{Esen K. Tütüncü}
\affiliation{%
  \institution{Institute of Neurosciences of the University of Barcelona}
  \city{}
  \country{Spain}}
\orcid{0000-0002-0050-0908}
\email{esenkucuktutuncu@ub.edu}

\author{Mar Gonzalez-Franco}
\affiliation{%
  \institution{Google}
  \city{Seattle}
  \country{USA}}
\email{margonzalezfranco@gmail.com}

\author{Eric J. Gonzalez}
\affiliation{%
 \institution{Google}
 \city{Seattle}
 \country{USA}}
\email{ejgonz@google.com}

\renewcommand{\shortauthors}{K. Tütüncü et al.}
\begin{abstract}
 We present HandOver, an extended reality (XR) interaction technique designed to unify the precision of traditional mouse input for object selection with the expressiveness of hand-tracking for object manipulation. With HandOver, the mouse is used to drive a depth-aware 3D cursor enabling precise and restful targeting — by hovering their hand over the mouse, the user can then seamlessly transition into direct 3D manipulation of the target object. In a formal user study, we compare HandOver against two ray-based techniques: traditional raycasting (Ray) and a hybrid method (Ray+Hand) in a 3D docking task. Results show HandOver yields lower task errors across all distances, and moreover improves interaction ergonomics as highlighted by a RULA posture analysis and self-reported measures (NASA-TLX). These findings illustrate the benefits of blending traditional precise input devices with the expressive gestural inputs afforded by hand-tracking in XR, leading to improved user comfort and task performance. This blended paradigm yields a unified workflow allowing users to leverage the best of each input modality as they interact in immersive environments.
\end{abstract}


\begin{teaserfigure}
  \includegraphics[width=1\textwidth]{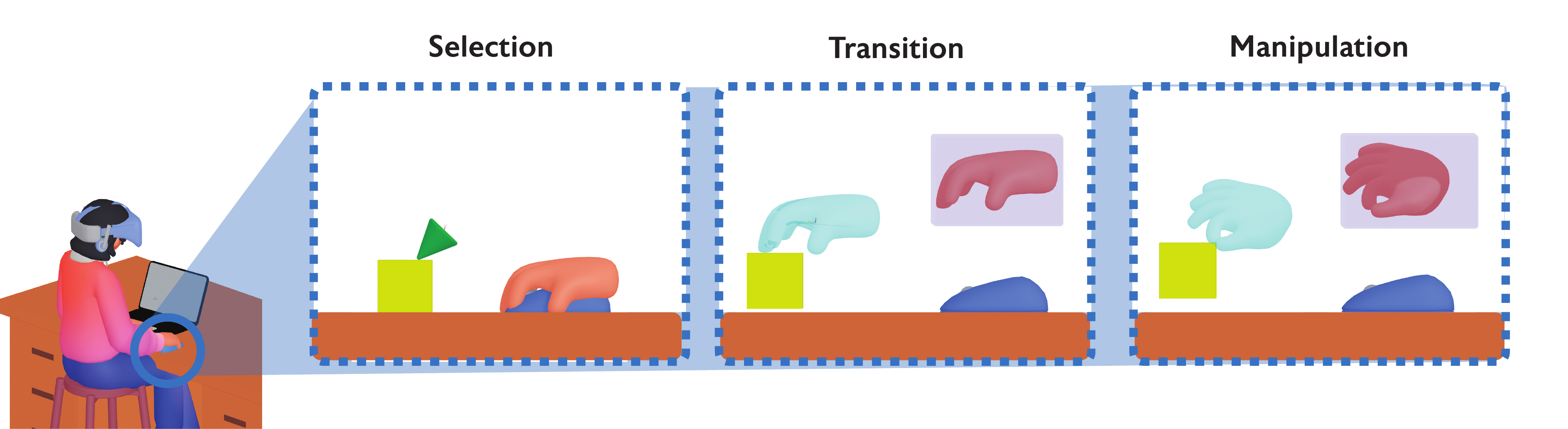}
    \centering
  \caption{The HandOver technique enables a smooth transition from mouse-based object selection to direct hand manipulation. Interaction stages: (Left) Selecting an object with the mouse, (Middle) Entering the handOver state by lifting the hand over the mouse, and (Right) Grasping the object with the hand.}
  \Description{Three different states of the Distant Hand Interaction}
  \label{fig:teaser}
\end{teaserfigure}


\maketitle
\section{Introduction}
Interaction research has long faced the challenge of combining the precision of traditional devices with the naturalness of spatial interaction. This challenge remains largely unsolved, creating an opportunity for novel approaches like the one we present in this paper. Classic mouse‐based input remains remarkably powerful for planar tasks, thanks to decades of refinements in input technologies and techniques \cite{hinckley2007input}, as well as foundational insights into human motor performance \cite{fitts1954information}. However, integrating mouse input into 3D environments presents fundamental challenges: primarily the difficulty of mapping 2D planar input to 3D spatial interactions, creating depth ambiguity, and managing proper visual feedback in immersive contexts \cite{fortune2017voronoi}. Previous approaches have attempted to solve these problems through various techniques, but no comprehensive solution has emerged \cite{nguyen2023hand,bergstrom2021evaluate}.

3D interaction can be better understood when divided into three distinct phases: selection, confirmation, and manipulation. Each phase presents unique challenges in spatial environments. The act of selecting and confirming an object often benefits from precise pointing or clicking—areas where mouse input excels. Yet once an object is “in hand,” direct control over its position and rotation is more natural with gesture-based 3D input or ray-casting techniques \cite{poupyrev1996go,pfeuffer2024design}. Even so, purely gestural approaches can cause tracking jitter, discomfort over long sessions, or confusion when users must isolate a small target amidst clutter \cite{tatzgern2016adaptive}. This tension suggests that no single approach has fully bridged the needs of precise selection, smooth confirmation, and fluid manipulation at varying distances.

Meanwhile, many authors have stressed that 3D interfaces have unique constraints not present in 2D. Reviews of target selection in virtual reality \cite{huang2019review} highlight this tension, pointing out that robust user performance typically demands a stable anchor, such as a mouse, for small precise targets, and more direct arm gestures for large or close ones. Some propose bridging these differences through fidelity models \cite{bonfert2024interaction} or “office of the future” approaches \cite{grubert2019office} that seamlessly integrate standard desktop gear into XR. Others argue that carefully balancing near‐field “mouse precision” with far‐field “arm extension” is crucial \cite{bowman1997evaluation}, as purely gestural solutions often amplify small tremors over distance \cite{mayer2018effect} or fail to provide the tactile anchoring a conventional mouse inherently offers.

To address this, we propose \textbf{HandOver}, a hybrid approach that integrates precise mouse-based selection with natural hand-based manipulation. HandOver combines a depth-aware cursor for far-field targeting with a virtual "hand clone" that projects natural gestures to distant objects. By leveraging mouse-based input for fine-grained targeting and using hand tracking for more intuitive manipulation, HandOver seeks to unify the user's sense of manual control from initial selection through final placement. The system dynamically transitions between these modalities so that small or closely spaced objects can be handled with mouse-level precision, while large or nearby objects can be manipulated via direct ``hands-on'' interaction. We conducted a user study to compare HandOver against two methods: Ray+Hand, a ray-based interaction technique that maintains consistent control-display gain regardless of distance, and Ray, the industry standard raycast interaction. We evaluated performance metrics such as targeting time, docking time, error, and physical workload across near, mid, and far distances, along with ergonomics. Our results shed light on how to optimize 3D interaction techniques by balancing familiarity (mouse-based selection) with the inherent manual expressiveness of gestural input, highlighting the hand’s foundational place in digital interfaces.
\smallskip
\smallskip

\section{Related Work}

\subsection{Mouse-Based Precision in 3D Interfaces}
The mouse remains one of the most precise input devices in computing, and its core strength—fine-grained control—has led to ongoing efforts to bring its benefits into 3D environments. Researchers have experimented with embedding mouse-like control into spatial interfaces, exploring how workflows from 2D design tools might translate into immersive settings. For instance, architectural tools have incorporated switchable 3D cursors to enable collaborative co-creation~\cite{dorta2016hyve}. Others have examined techniques for toggling between absolute and relative pointing, revealing how traditional input can remain effective in multi-dimensional tasks~\cite{forlines2006hybridpointing}. More recently, desktop mouse input has been reintroduced into VR, where it has shown advantages over mid-air gestures, particularly in tasks that demand delicate alignment~\cite{zhou2022depth}.
\smallskip
\smallskip

\subsection{Precision-Enhancing Pointing Techniques}
Beyond mouse emulation, a number of approaches aim to preserve or improve pointing accuracy in spatial settings through dynamic adaptation. Modulating the control-display gain based on target distance or size has been shown to support more precise selection~\cite{konig2009adaptive}. In augmented reality, head- and eye-based methods have been evaluated for their ability to replicate the targeting fidelity of mouse input~\cite{kyto2018pinpointing}. However, precision degrades rapidly when users move away from the body or toward large displays. Studies of distant freehand pointing suggest that while the mouse excels in near-field contexts, it often requires complementary methods to maintain accuracy at scale~\cite{vogel2005distant}.
\smallskip
\smallskip

\subsection{Ray-Based Selection and Its Limitations}
Ray-based interaction is one of the most common strategies for selecting distant objects in VR. This technique, often implemented through controllers or hand gestures, maps small movements to large cursor displacements, making it effective for long-range tasks. However, studies have shown that this approach suffers from instability, including tremor amplification and arm fatigue, especially when users hold their hands at full extension for prolonged periods~\cite{rantamaa2023comparison, kim2023perspective}. Various refinements have been proposed to improve accuracy, including ray widgets, gesture-based rays, and predictive models that incorporate head movement or user posture~\cite{lee2003evaluation, zhao2024novel, baloup2019raycursor, henrikson2020head}, but the underlying physical constraints often remain.
\smallskip
\smallskip

\subsection{Adaptive and Assistive Targeting Models}
To offset the ergonomic and stability issues of direct pointing, assistive selection techniques have been developed. Many of these focus on adjusting the selection radius or cursor behavior based on task context. For example, semantic and dynamic control-display gain adaptations have been shown to improve target acquisition times without increasing physical effort~\cite{teather2011pointing, grossman2005bubble}. Other systems address common problems such as cursor drift and depth ambiguity, using predictive modeling, stereo visualizations, or corrective overlays~\cite{mayer2015modeling, ramcharitar2018ezcursorvr, teather2013pointing}. Depth-aware hand gestures and viewpoint adaptation further extend these strategies to support more robust input in immersive 3D environments~\cite{kim2022viewfindervr, wang2024cursor}.
\smallskip
\smallskip

\subsection{Technique Comparisons and Ergonomics}
Comparative studies have played a key role in evaluating how input methods scale across tasks, distances, and devices. Techniques such as nonlinear reach extension~\cite{poupyrev1996go}, pen versus controller input~\cite{pham2019pen}, and ray length variation~\cite{batmaz2020effect} have been tested for their effectiveness and precision. Multi-display environments have motivated new cursor models like the Perspective Cursor~\cite{nacenta2006perspective}, designed to support fluid transitions across screens. Meanwhile, work on cross-device orchestration ~\cite{brudy2020surfacefleet} has emphasized the importance of consistent input metaphors, even if not directly focused on pointing.

In parallel, other research has examined the physical and cognitive load imposed by immersive interaction. Studies have explored offset calibration~\cite{mayer2018effect}, posture-induced drift~\cite{myers2002interacting}, and full-body fatigue during repetitive spatial tasks~\cite{montano2017erg, petford2018pointing, sun2018comparing}. These findings underline the need to evaluate both performance metrics and embodied cost when comparing spatial input methods.
\smallskip
\smallskip

\subsection{Hybrid Input Devices and Multimodal Transitions}
Prior work has explored hybrid devices that blend surface-based precision with mid-air flexibility. The 3D Gesture Mouse \cite{7460060} and Roly-Poly Mouse \cite{10.1145/2702123.2702244} integrated tangible form factors to enable transitions between desktop and spatial control. Systems like SurfAirs and AirMouse \cite{10.1145/3544548.3580877, 10.1007/978-3-642-03658-3_28} emphasized seamless input switching between modalities, using either custom hardware or motion sensing to maintain continuity.

Our work contributes to this line by offering a hybrid technique that avoids explicit mode switching and does not require additional devices. \textit{HandOver} leverages a unified input model where cursor-based targeting and hand-based manipulation are part of a continuous workflow. Inspired by emerging trends in cross-device and proxemic interaction \cite{10536201, 10.1145/3613904.3642323}, it builds on the design space of "in-between" techniques highlighted by recent multimodal toolkits.

HandOver is situated within this broader trajectory as a seated, XR-native technique that supports high-fidelity precision and embodied alignment in the same interaction sequence, broadening the scope of hybrid interaction design.

\smallskip
\smallskip

\section{System Design}
\subsection{Depth-Adaptive Cursor}
The \emph{depth-adaptive} 3D cursor enhances target selection in XR by adapting the cursor’s depth based on its position and the surrounding objects. Implemented based on the work by Zhou et al. \cite{zhou2022depth}, the system works by first calculating a screen-space update using the hardware mouse’s 2D delta, then projecting a ray from the camera to detect the nearest surface within a maximum depth. If no collision occurs, depth estimation is done using a Voronoi diagram blending nearby scene colliders, ensuring smooth depth transitions. The cursor depth is then smoothed to avoid abrupt jumps, and its size is rescaled according to its distance from the camera to maintain a consistent visual appearance. This technique improves both performance and comfort in XR environments by providing a stable, responsive cursor behavior.

\subsection{Mouse-Hand Transition}
Alongside mouse control, the system monitors the user’s physical hand position. An invisible “hover bounding box,” placed approximately $8\,\mathrm{cm}$ above the real hand’s anchor point (taking into account the height of the mouse and the hand placed over it), allows the system to detect when the user transitions from a purely mouse‐driven approach to a more direct hand interaction. Specifically, once the hand enters this box (within a vertical offset of $\approx 0.05\,\mathrm{m}$ above the anchor), and the cursor remains stationary below a $1\,\mathrm{mm}$ threshold across multiple frames, the application interprets this as intent to manipulate objects in 3D.

At that moment, the 3D cursor becomes temporarily hidden, while a \emph{hand clone} spawns at the cursor’s last known position. The cloned hand is a copy of the user’s original hand model but is augmented to reduce fatigue. A configurable gain factor amplifies small real-hand movements into larger virtual displacements, aiding far-field manipulation. The displacement between the real hand and its clone is clamped to a maximum offset of $4.0\,\mathrm{m}$ to avoid unintended drift.

\begin{figure}[htbp]
  \centering
  \includegraphics[width=\columnwidth]{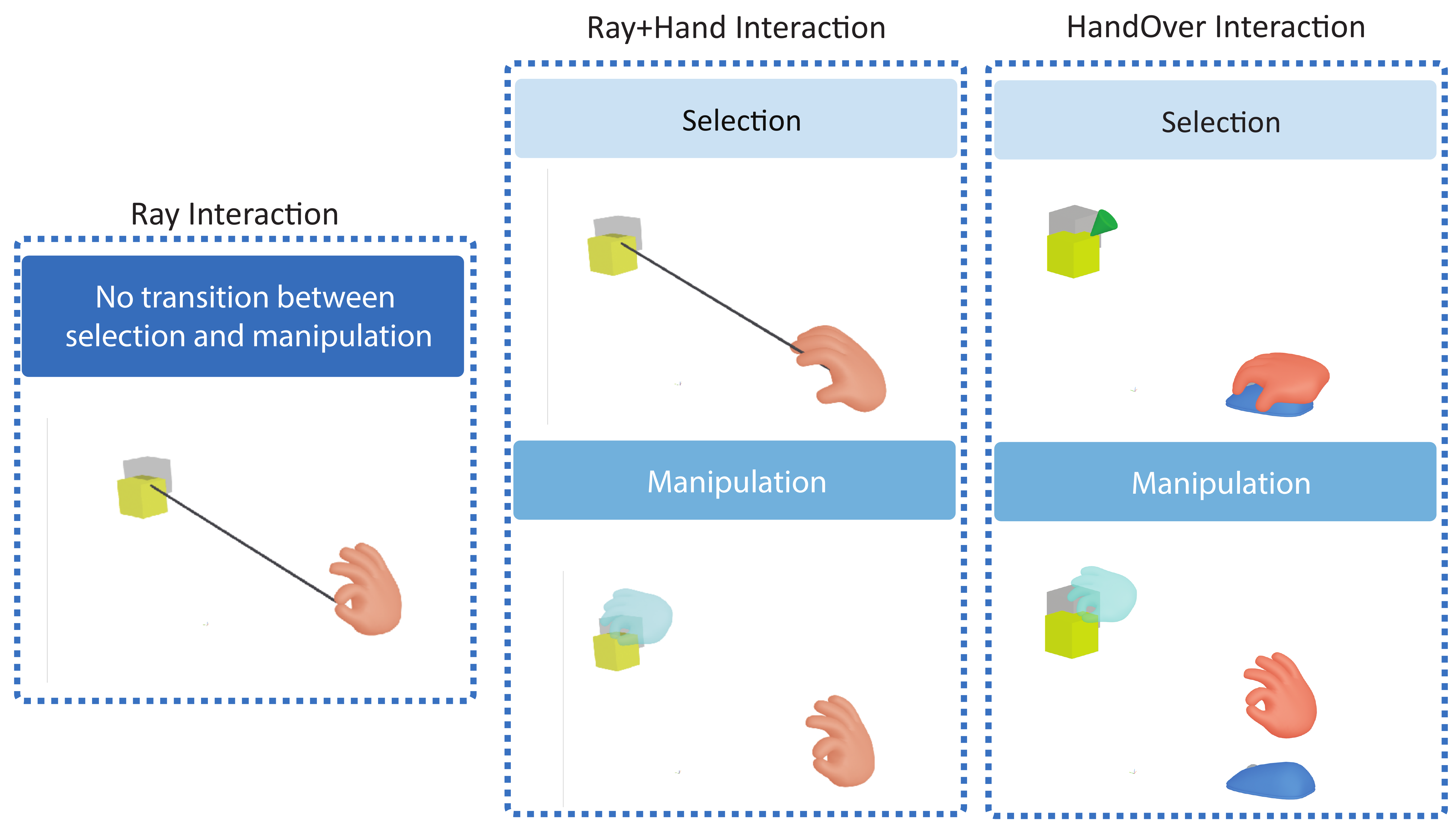}
  \caption{Comparing the three interaction techniques Ray, Ray+Hand, and HandOver, across the selection-to-manipulation transition.}
  \Description{Comparing Ray, Ray+Hand, and HandOver in the transition from selection to manipulation.}
  \label{fig:interactions}
\end{figure}

Once active, the clone checks for overlapping objects within a $5\,\mathrm{cm}$ radius. If the user performs a pinch gesture (index and thumb pinch strength $> 0.5$), the system attaches the overlapping object to the clone’s transform. The object remains “grabbed” until the pinch ends. For a hand to be considered “closed,” the pinch strength reported by the headset’s hand‐tracking API must exceed $0.3$–$0.5$ across multiple fingers (e.g., thumb and index for precision grip, or all fingers for a fist). Conversely, an “open palm” is defined by all finger pinch strengths falling below $0.05$, along with a palm‐facing‐camera check where the dot product between the palm normal and the vector to the camera is less than $-0.90$.

A fade-in/fade-out mechanism (duration $0.2\,\mathrm{s}$) governs the transition between interaction modes, providing immediate visual feedback. When the above activation conditions are met, the system fades out the 3D cursor and fades in the hand clone. During clone activity, mouse input is ignored to avoid conflict. If the user releases the pinch and moves their hand back toward the original anchor (within $\approx 2\,\mathrm{cm}$), the system reverses the transition: the clone fades out, the cursor reappears at its prior position, and mouse control resumes.

This mechanism ensures that only one interaction modality is functionally active at a time, either the hand clone or the 3D cursor—based on real-time behavioral cues. This mutual exclusivity preserves clarity in user input and avoids ambiguity in object manipulation.

\section{Study Design}

\subsection{Task Design}

The user study follows Fitts’ Law~\cite{fitts1954information} principles to evaluate trade-offs between speed and accuracy in virtual object placement. Three distinct distances (2\,m, 4\,m, and 6\,m) simulate near, mid, and far interactions. 

To ensure angular variety and consistent target spacing, we positioned objects using a circular layout around the participant. The $360^\circ$ circle was divided into eleven equal increments ($\approx 32.7^\circ$), and a randomized index was selected as a starting point. For each trial, the next target angle was determined by skipping halfway around the circle, maximizing spatial separation and minimizing learning or habituation effects. This layout mimics classic Fitts-style amplitude and angle separation.

Targets were placed at a radius of $\approx 1\,\mathrm{m}$ from the center, which was aligned with the user’s eye height at study start (camera’s $y$ position). This height anchoring ensured ecological validity by simulating seated interaction scenarios typical in desktop XR setups. The $z$-coordinate of the circular center was set according to the active trial distance (near, mid, or far). 

Each square target was accompanied by a corresponding interactable object, spawned with an offset of up to $\pm 0.2\,\mathrm{m}$ in a random direction around the target. This slight spatial mismatch was intentional, simulating natural placement imprecision and requiring users to make minor spatial corrections. Participants were instructed to grab or maneuver the object into the target’s bounding zone with minimal error. Once satisfied, they pressed the space bar to confirm placement, destroying the current trial objects and advancing to the next. This spacebar-triggered advancement was chosen to prevent accidental dragging effects that sometimes occurred on pinch release.

After 11 placements at a given distance, the study transitioned to the next. All three distances were tested per interaction technique, resulting in nine condition blocks (3~distances~$\times$~3~techniques). The order of blocks was counterbalanced using a Latin square design.

\begin{figure}[h]
  \centering
  \includegraphics[width=\columnwidth]{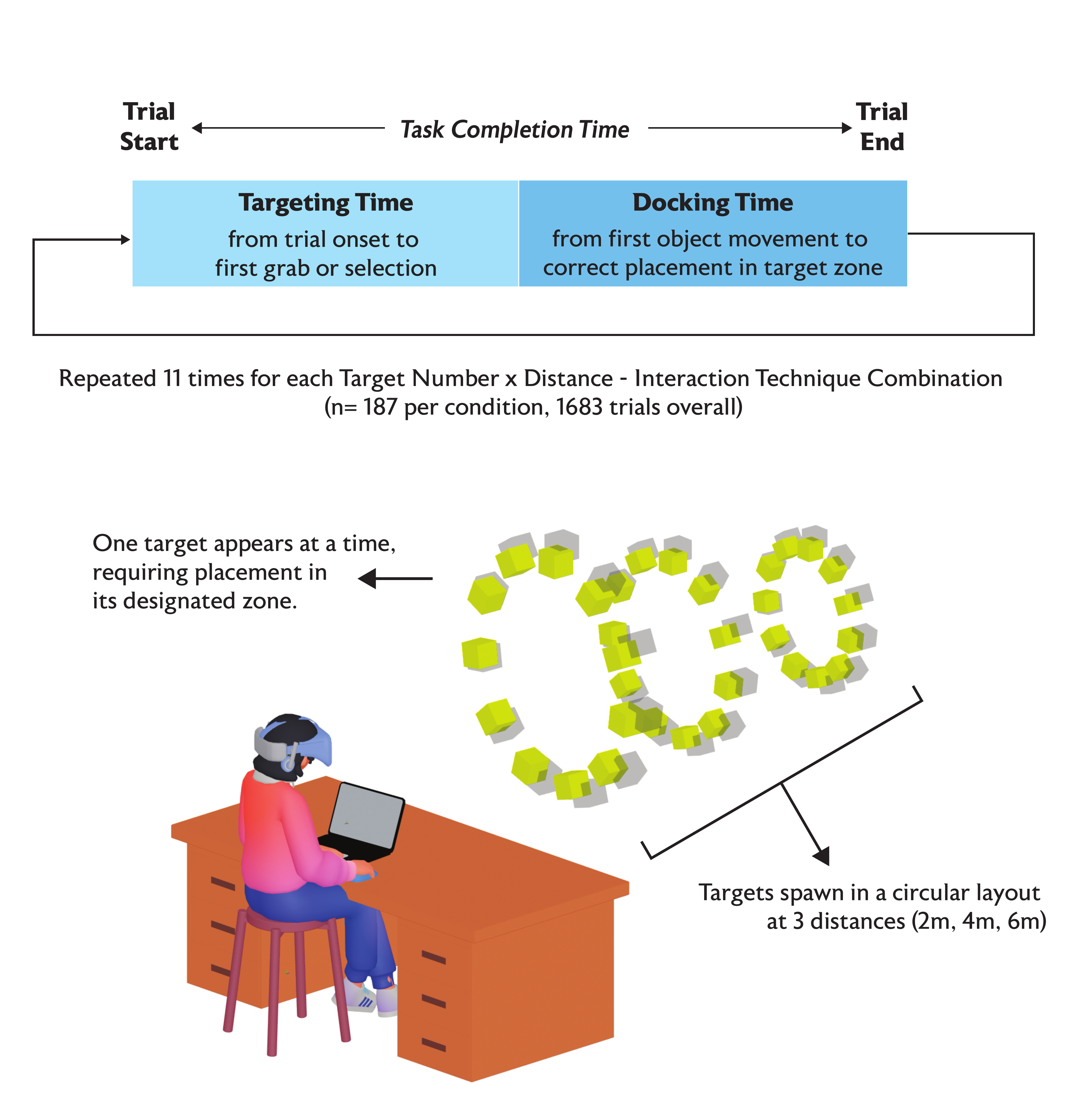}
  \caption{Experimental design showing the circular target layout and interaction phases. Users perform targeting (from trial onset to first grab/selection) and docking (from object movement to final placement). The trial repeats across distances and techniques.}
  \label{fig:task_design}
\end{figure}

We compare three methods for remote object manipulation. 
\emph{HandOver} is primarily mouse‐based: a standard 2D mouse drives a 3D cursor, and if the cursor becomes stationary while the user’s real hand hovers close by, the system spawns a virtual “hand clone” at the cursor’s location for direct pinch gestures. 
\emph{Ray+Hand} uses a ray from the user’s real hand to the target, then spawns a hand clone at the ray’s collision point upon pinch detection. The CD gain is dynamically scaled to the user’s arm motion based on distance, preserving consistent control at both near and far placements by mimicking 1:1 movement. 
By contrast, \emph{Ray} implements a purely geometric relationship, so as distance grows, small real‐arm movements become magnified. Minor deviations thus yield large cursor shifts at far ranges, replicating the current ray-based interaction techniques.

\smallskip
\smallskip

\subsection{Apparatus \& Implementation}
All interactions were developed in Unity 2022.3.39f1, running on a portable laptop connected to an Oculus Quest~3 headset. The system continuously monitors cursor coordinates, ray intersections, and pinch strengths via the Oculus hand-tracking API. Logged data includes object coordinates, user hand positions, gesture states, and interaction events (e.g., grabs, releases). All three conditions run within the same Unity project and share foundational code to ensure consistency, differing only in how each interprets the user’s hand movements or mouse inputs.

In addition, we implemented a simple state machine to handle both the active interaction technique (HandOver, Ray+Hand, or Ray) and the distance condition (2\,m, 4\,m, or 6\,m). At the start of each trial, the state machine updates the system’s parameters—such as where to spawn objects or whether to enable mouse-based transitions, and ensures each interaction mode’s logic is correctly activated or deactivated. This approach streamlines the code responsible for switching techniques, labeling trial data, and orchestrating transitions between near, mid, and far tasks.
\smallskip
\smallskip

\subsection{Participants}
An in-person study was conducted with 17 participants (9 male, 8 female) with a mean age of 28.94 years (SD = 4.37). Participants reported moderate familiarity with VR usage (M = 4.06, SD = 1.89) and slightly less experience with VR development (M = 3.24, SD = 2.02) on a 7-point Likert scale. Participants were compensated for their time, with no incentives linked to performance. The institution’s review board approved the study.

\begin{table*}[t]
\centering
\small
\setlength{\tabcolsep}{6pt}
\renewcommand{\arraystretch}{1.5}

\begin{tabular}{>{\raggedright\arraybackslash}p{2.5cm} 
                >{\centering\arraybackslash}p{2.5cm} 
                >{\centering\arraybackslash}p{1.5cm} 
                >{\centering\arraybackslash}p{1.5cm} 
                >{\raggedright\arraybackslash}p{4cm}}
\toprule
\textbf{Metric} & \textbf{HandOver (HO)} & \textbf{Ray+Hand (R+H)} & \textbf{Ray (R)} & \textbf{Significance Statistics} \\
\midrule

\textbf{Targeting Time (s)} 
& 5.3 (±0.63) 
& 4.1 (±0.79) 
& 6.4 (±0.68)
& \makecell[l]{
Technique: $p < .001$,  $\eta^2 = .021$; \\
Distance: $p < .001$,  $\eta^2 = .038$ \\
Interaction: $p = .366$}
\\

\addlinespace

\textbf{Docking Time (s)} 
& 5.1 (±0.58) 
& 4.2 (±0.79) 
& 6.9 (±0.64)
& \makecell[l]{Technique: $p < .001$, $\eta^2 = .047$;\\
Distance: $p < .001$, $\eta^2 = .004$ \\
Interaction:
$p = .003$, $\eta^2 = .013$} \\

\addlinespace

\textbf{Docking Error (cm)} 
& 2.3 (±.29) 
& 2.6 (±.2) 
& 3.7 (±.35)
& \makecell[l]{Technique: $p < .001$, $\eta^2 = .112$;\\
Distance: $p < .001$, $\eta^2 = .042$ \\
Interaction: $p = .02$, $\eta^2 = .011$ 
} \\

\addlinespace

\textbf{RULA Score (composite)} 
& 5.8 (±.19)
& 6.8 (±.21) 
& 7.2 (±.20) 
& \makecell[l]{Technique: $p < .001$;
Distance: n.s.; \\
Interaction: $p = .62$; $\eta^2 = .112$} \\

\addlinespace

\textbf{Accumulated Movement (cm)} 
& 956.4 (±18.6) 
& 1085.9 (±23.6) 
& 1585.5 (±40.7) 
& \makecell[l]{
Technique: $p < .001$; 
Distance: n.s.; \\
Interaction: $p = .235$; $\eta^2 = .053$}\\

\bottomrule
\end{tabular}
\caption{Comparative results across interaction techniques. Values represent means aggregated across all distances, with standard error of the mean (SEM) in parentheses. Lower scores indicate better performance across all metrics.}
\label{tab:comparative_results}
\end{table*}

\smallskip
\smallskip

\section{Results}
We report results across multiple task-level and ergonomic metrics, using statistical tests appropriate to each metric’s scale and distribution. For normally distributed, continuous variables (e.g., targeting time, docking time, and docking error), we conducted two-way repeated-measures ANOVAs with technique and distance as within-subject factors, followed by pairwise Tukey-corrected comparisons. For ordinal or non-normally distributed data—such as RULA scores, accumulated joint movement, and NASA-TLX subscales—we applied Kruskal–Wallis tests with Wilcoxon signed-rank post hoc tests using Holm correction. Categorical data such as the number of regrabs were analyzed using the Friedman test with post-hoc Wilcoxon comparisons. Effect sizes and corrected p-values are reported where applicable. Significance levels: $^{*}p < 0.05$, $^{**}p < 0.01$, $^{***}p < 0.001$.

A summary of mean values and significance statistics aggregated across all distances is provided in Table~\ref{tab:comparative_results}. The subsections that follow present detailed breakdowns by distance and condition.

\subsection{Overall Performance Metrics}
For each condition, the number of trials was \(n = 187\), with a total trial count  of \(N = 1683\). The degrees of freedom were as follows: for interaction technique, \(df = 2\); for distance, \(df = 2\); for the interaction effect (Technique × Distance), \(df = 4\); and for residuals, \(df = 1675\).

\subsubsection{Targeting Time}
\begin{figure*}[t]
  \centering
  \includegraphics[width=2.1\columnwidth]{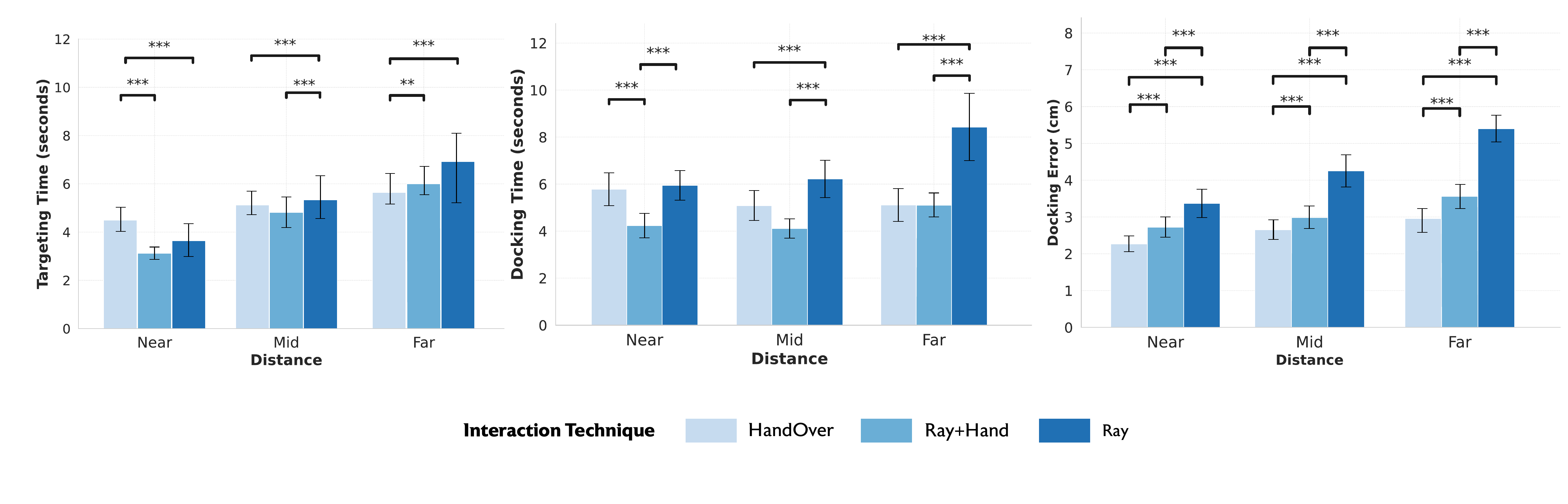}
  \caption{Performance comparison of the three interaction techniques (\emph{HandOver}, \emph{Ray+Hand}, and \emph{Ray}) across three distances (Near, Mid, Far) for (left)~targeting time, (center)~docking time, and (right)~docking error. Bars show means with error bars denoting standard error of the mean.}
  \label{fig:joint_vis}
\end{figure*}

We distinguish \textit{targeting time} (from trial onset to first grab or selection) from the overall task completion time.  
The first plot in Figure~\ref{fig:joint_vis} presents targeting durations across different distances (Near, Mid, Far).

There were significant main effects of both technique ($p < 0.001$, $\eta^2 = 0.021$) and distance ($p < 0.001$, $\eta^2 = 0.038$), but no significant interaction between the two factors ($p = 0.366$, $\eta^2 = 0.004$).

Targeting performance varied significantly across interaction techniques at all distances. At Near distance, \textit{Ray+Hand} achieved the fastest targeting times, significantly outperforming \textit{HandOver} ($U = 16408.5$, $p < 0.001$, $d = 0.290$), while no significant difference was found between \textit{Ray+Hand} and \textit{Ray} ($U = 10369.5$, $p = 0.261$, $d = 0.013$).  

At Mid distance, \textit{Ray+Hand} performed significantly better than \textit{Ray} ($U = 14199.5$, $p < 0.001$, $d = 0.324$), with no significant difference compared to \textit{HandOver}.

At Far distance, \textit{HandOver} resulted in the shortest targeting times, significantly faster than both \textit{Ray+Hand} ($U = 12879.0$, $p < 0.001$, $d = 0.312$) and \textit{Ray} ($U = 12298.5$, $p = 0.004$, $d = -0.238$, where the negative sign simply indicates HandOver's mean time was lower), which did not significantly differ from one another ($U = 10511.0$, $p = 0.611$)..

\subsubsection{Docking Time}
\textit{Docking time} refers to the interval between the first object movement and the successful placement of the object in the target zone.  
The second plot in Figure~\ref{fig:joint_vis} shows docking times across interaction techniques and distances (Near, Mid, Far).

Docking performance varied significantly across both techniques ($p < 0.001$, $\eta^2 = 0.047$) and distances ($p < 0.001$, $\eta^2 = 0.004$), with a significant interaction between the two factors ($p = 0.003$, $\eta^2 = 0.013$). Across all distances, \textit{Ray} consistently required more time to complete the docking phase than both \textit{HandOver} and \textit{Ray+Hand}.

At Near distance, \textit{Ray+Hand} achieved the shortest docking times, significantly outperforming both \textit{HandOver} ($U = 13261$, $p = 0.001$, $d = 0.312$) and \textit{Ray} ($U = 7645$, $p < 0.001$, $d = -0.551$). No significant difference was found between \textit{HandOver} and \textit{Ray} at this distance ($U = 10024.5$, $p = 0.329$).

At Mid distance, both \textit{HandOver} and \textit{Ray+Hand} outperformed \textit{Ray}, with significant differences observed for both comparisons (\textit{HandOver} vs \textit{Ray}: $U = 8248$, $p = 0.003$; \textit{Ray+Hand} vs \textit{Ray}: $U = 7491$, $p < 0.001$). No significant difference was found between \textit{HandOver} and \textit{Ray+Hand} ($U = 11413$, $p = 0.159$).

At Far distance, \textit{Ray} again resulted in significantly longer docking times than both \textit{HandOver} ($U = 7255$, $p < 0.001$) and \textit{Ray+Hand} ($U = 6851$, $p < 0.001$), while \textit{HandOver} and \textit{Ray+Hand} did not differ significantly from each other ($U = 10978.5$, $p = 0.374$).

These results highlight that while \textit{Ray} consistently imposed the greatest time cost for docking, \textit{Ray+Hand} demonstrated a clear advantage at Near distance, and \textit{HandOver} maintained stable performance across all distances.

\subsubsection{Docking Error}
The third plot in Figure~\ref{fig:joint_vis} shows \textit{Docking Error} (cm), defined as the Euclidean distance between the object’s final position and the center of the target prefab. \textit{HandOver} consistently achieved the lowest error, ranging from 2.3\,cm at Near to 2.9\,cm at Far. \textit{Ray+Hand} produced moderately higher errors (2.6--3.2\,cm), while \textit{Ray} resulted in the largest errors, often exceeding 3.5\,cm—highlighting the difficulty of precise alignment at extended reach.

There were significant main effects of both \textit{Technique} and \textit{Distance} ($p < 0.001$), as well as a significant interaction between them ($p = 0.002$), indicating that technique performance varied with distance. The effect size for \textit{Technique} was substantial ($\eta^2 = 0.116$), reinforcing the importance of interaction method in determining placement precision.

Post-hoc comparisons confirmed that \textit{HandOver} produced significantly lower docking errors than both \textit{Ray+Hand} and \textit{Ray} at all distances ($p < 0.001$). While \textit{Ray+Hand} outperformed \textit{Ray} across Near, Mid, and Far ($p < 0.001$), its error rates remained noticeably higher than \textit{HandOver}. These results reinforce the advantage of \textit{HandOver} for precise object placement, particularly in scenarios where accuracy is critical across varying interaction distances.

\subsubsection{Regrab Counts}
Regrab counts, which reflect how often participants released and re-grabbed the same object within a trial, varied notably by technique. These values represent average counts per trial, aggregated across participants. \textit{Ray} showed the highest frequency (\(\approx 1.4\) at Far), indicating that users often had to adjust their grasp more than once per trial. \textit{HandOver} remained lower (0.7--0.8), while \textit{Ray+Hand} ranged from 0.5 (Mid) to 1.0 (Far). Technique had a significant effect on regrab rates (\(p < 0.01\)), with post-hoc comparisons indicating that \textit{Ray} led to significantly more regrabs than \textit{HandOver} (\(p < 0.05\)).

\begin{figure}[t]
  \centering
  \includegraphics[width=1\columnwidth]{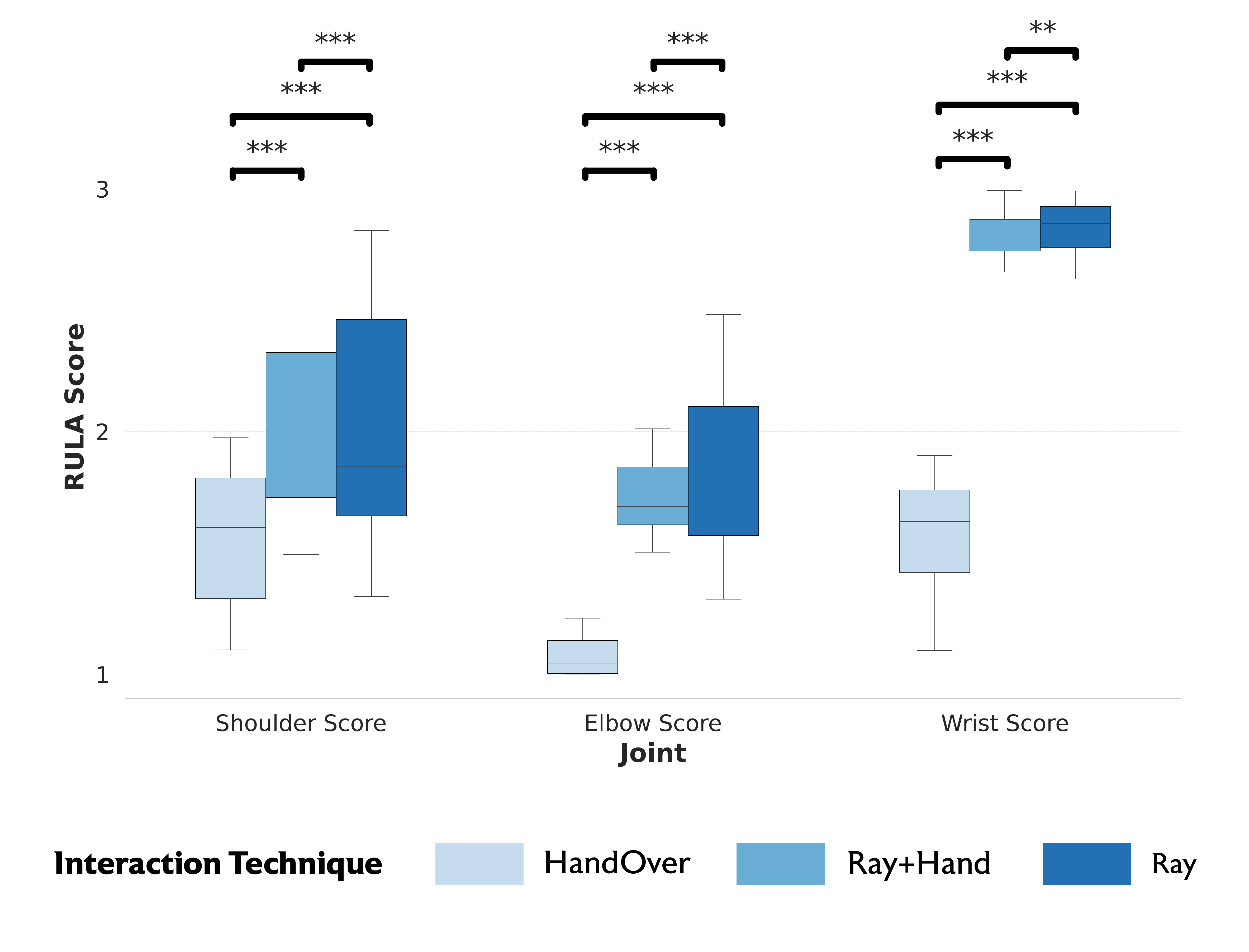}
  \caption{Per‐joint RULA Scores (1=Low, 2=Moderate, 3=High) for 
  shoulder, elbow, and wrist flexion across each technique.}
  \label{fig:rula}
\end{figure}

\begin{figure}[t]
  \centering
  \includegraphics[width=\columnwidth]{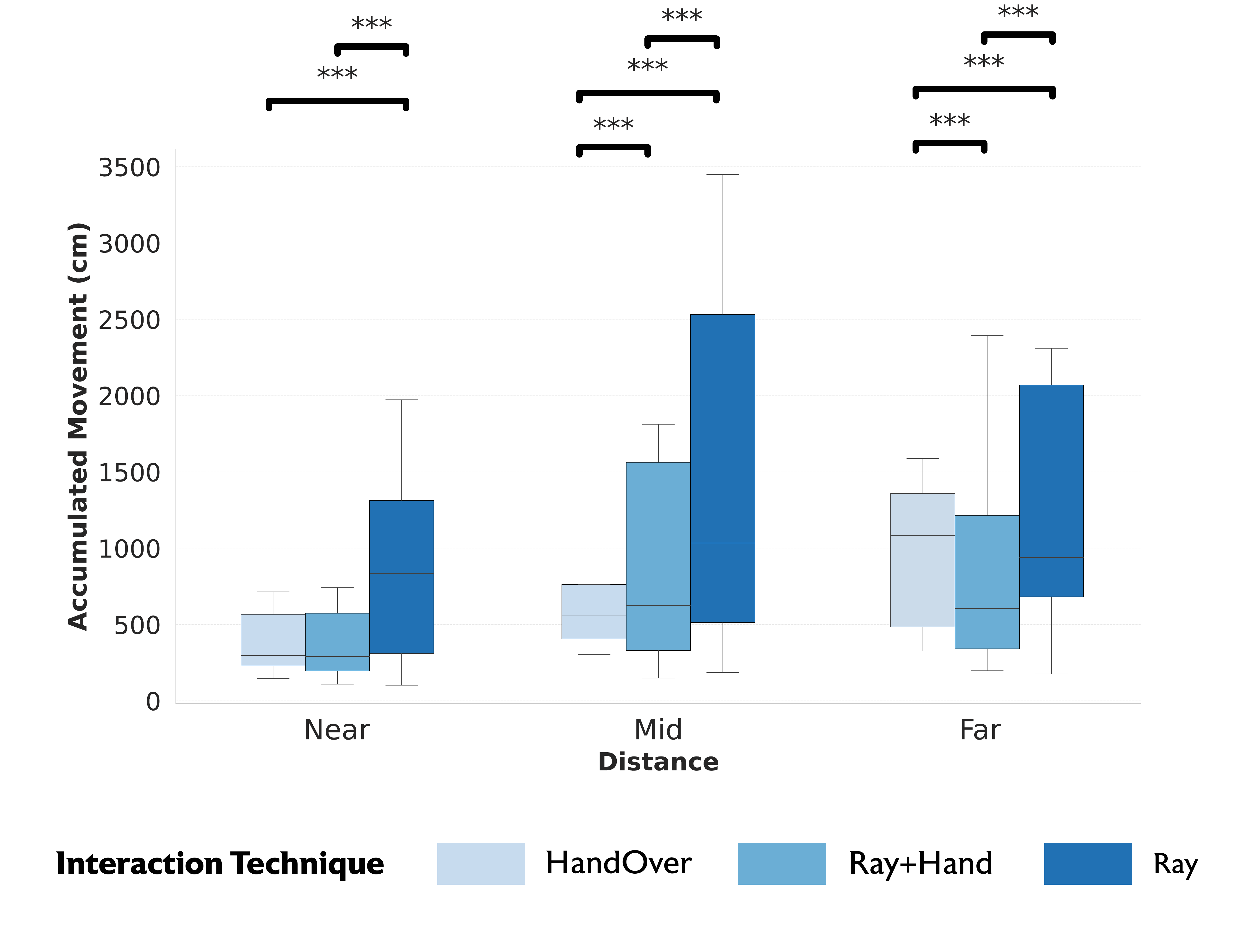}
  \caption{Accumulated movement (sum of joint displacements) for shoulder, elbow, and wrist across interaction techniques and distances. }
  \label{fig:accumulated_movement}
\end{figure}

\smallskip
\smallskip
\subsection{Movement Metrics}
\smallskip
\smallskip
\subsubsection{RULA Posture Analysis}
We employed a modified Rapid Upper Limb Assessment (RULA) to evaluate ergonomic strain across the wrist, elbow, and shoulder joints. Following standard guidelines~\cite{mcatamney1993rula}, joint angles were computed from tracked 3D positions using vector-based dot-product calculations. For each frame, the angle \(\theta\) between relevant joints (e.g., shoulder--elbow, elbow--wrist) was computed as:

\[
\theta = \text{arctan2}(v_z, v_y),
\]

where \(v_y\) and \(v_z\) are the components of the limb vector in the vertical and depth directions. Postural strain was then assessed based on predefined ergonomic thresholds (e.g., 20\textdegree{} and 45\textdegree{} for the shoulder), with higher scores indicating greater deviation from a neutral posture.


To ensure a fair ergonomic comparison, we applied a 1-point reduction to all RULA scores in the \textit{HandOver} condition. This adjustment reflects the reduced muscular load and joint deviation observed during wrist-supported input postures in seated configurations, as recommended in ergonomics literature \cite{mcatamney1993rula}. Final scores were averaged across distances for each technique to provide a robust comparison (Figure \ref{fig:rula}).

Interaction technique had a significant effect on overall RULA scores (\(H = 16.79\), \(p = 0.0002\)), with a moderate effect size (\(\eta^2 = 0.112\)). Post-hoc comparisons confirmed that \textit{HandOver} resulted in significantly lower ergonomic strain than both \textit{Ray+Hand} (\(U = 1464.00\), \(p = 0.0003\), \(d = 0.446\)) and \textit{Ray} (\(U = 1437.00\), \(p = 0.0006\), \(d = 0.419\)). No significant difference was observed between \textit{Ray+Hand} and \textit{Ray} (\(U = 976.00\), \(p = 0.7714\), \(d = -0.036\)).

\begin{figure*}[t]
  \centering
  \includegraphics[width=2\columnwidth]{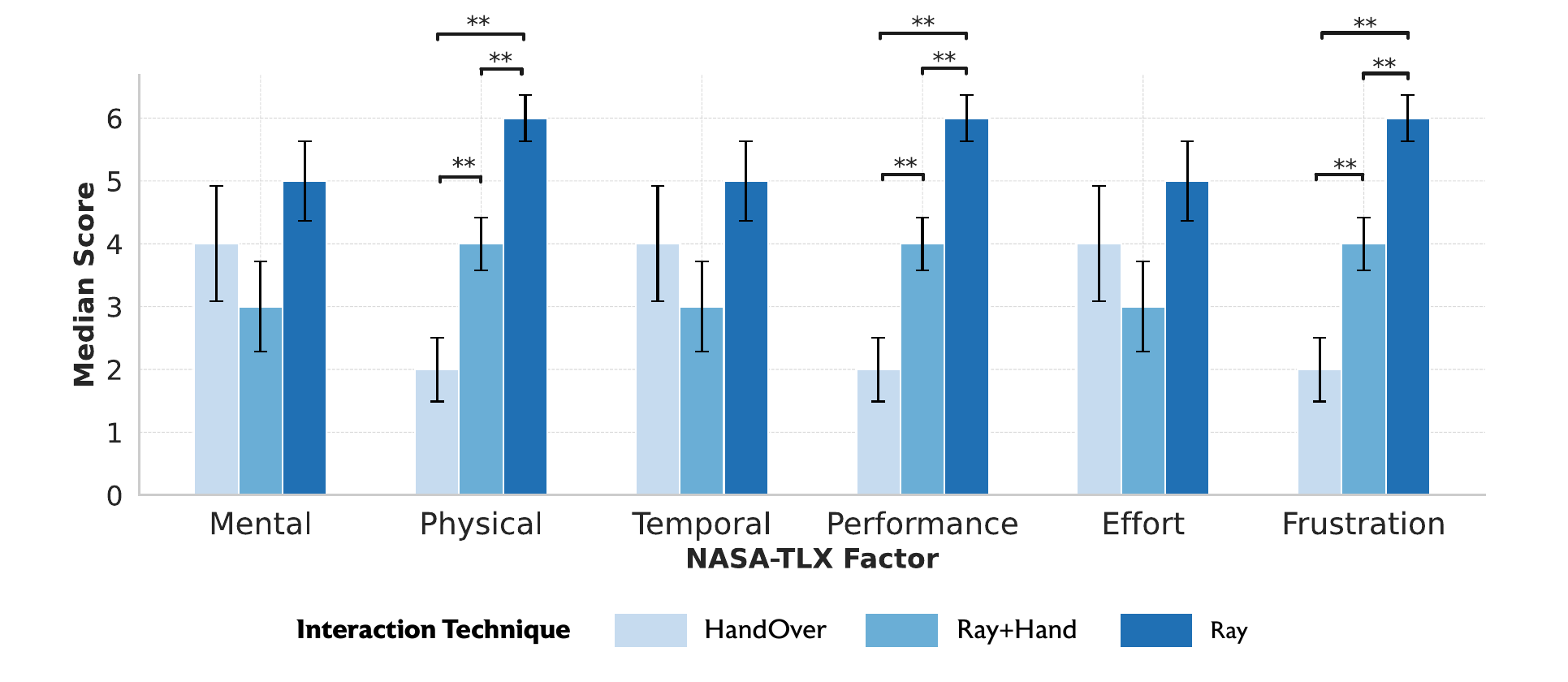}
  \caption{Mean NASA-TLX Ratings for each interaction technique (HandOver, Ray+Hand, Ray) on six subscales.}
  \label{fig:nasatlx}
\end{figure*}

\smallskip
\smallskip
Across joints, \textit{Ray} produced the highest RULA scores, particularly in the shoulder and wrist, with over 55\% of frames in non-neutral shoulder postures at Far distance. Wrist strain also peaked with \textit{Ray}, exceeding 60\% non-neutral duration. In contrast, \textit{HandOver} maintained lower RULA scores across joints, reflecting more relaxed postures. \textit{Ray+Hand} remained intermediate, slightly improving over \textit{Ray} in elbow alignment but still resulting in elevated wrist strain.

Summing across joints and muscle-load increments, \textit{Ray} scored highest overall (Far: \(M = 7.2\), \(SEM = 0.20\); Mid: \(M = 6.9\), \(SEM = 0.18\)). \textit{Ray+Hand} followed closely (Far: \(M = 6.8\), \(SEM = 0.21\)), while \textit{HandOver} yielded significantly lower strain levels (Far: \(M = 5.8\), \(SEM = 0.19\); Near: \(M = 5.6\), \(SEM = 0.17\)).

\smallskip
\smallskip

\subsubsection{Accumulated Movement}
To quantify physical effort, we computed \emph{accumulated movement} as the sum of 3D displacements for the shoulder, elbow, and wrist joints across all frames:
\[
  \sum_{t=1}^{N} \bigl\|\mathbf{P}_t - \mathbf{P}_{t-1}\bigr\|,
\]
where \(\mathbf{P}_t\) denotes the tracked joint position at time \(t\).

Interaction technique had a statistically significant effect on accumulated movement ($H = 91.60$, $p < 0.001$), with a small-to-moderate effect size ($\eta^2 = 0.053$). Post-hoc comparisons revealed significant pairwise differences between all techniques ($p < 0.001$). \textit{Ray} consistently resulted in the greatest physical effort, particularly at Mid ($M = 1583.53$, SEM = $\pm$843.58) and Far distances ($M = 1253.60$, SEM = $\pm$936.14).

Pairwise analyses showed that the difference between \textit{Ray} and \textit{HandOver} was largest at Mid ($d = -0.333$) and Far ($d = -0.155$), supporting the notion that direct ray-casting required additional arm repositioning and fine-tuning at moderate-to-long distances. At Near distances, \textit{HandOver} and \textit{Ray+Hand} were not meaningfully different ($d = 0.024$), but both required significantly less movement than \textit{Ray} ($p < 0.001$), reinforcing the efficiency of hybrid and direct hand interactions in close-range targeting.

Despite relatively small standard error of the mean (SEM) values—indicating consistent within-group behavior—the differences in accumulated movement between techniques remained statistically significant.

\smallskip
\smallskip
\subsubsection{NASA-TLX Workload Ratings and Subjective Feedback}
We used the NASA-TLX questionnaire \cite{hart1988development} to assess subjective workload across six dimensions: \textit{Mental Demand}, \textit{Physical Demand}, \textit{Temporal Demand}, \textit{Performance}, \textit{Effort}, and \textit{Frustration}.  
Figure~\ref{fig:nasatlx} presents the \emph{median} scores and \emph{standard error of the mean (SEM)} for each subscale across the three techniques.

Among the six subscales, differences were most pronounced in \textit{Physical Demand} ($p = 0.0053$), \textit{Performance} ($p = 0.0057$), and \textit{Frustration} ($p = 0.0065$), highlighting meaningful differences in perceived effort, effectiveness, and user comfort.

\textit{HandOver} received medians of 2.0 in both \textit{Physical Demand} and \textit{Frustration} (SEM = 0.51)—and was also rated highest in \textit{Performance} (median = 2.0). This suggests that participants found it both comfortable and effective to use. In contrast, \textit{Ray} was rated as the most physically demanding (median = 6.0, SEM = 0.37), with equally high levels of \textit{Frustration} and the lowest perceived \textit{Performance} (also median = 6.0). Participants often noted that sustaining ray control over distance led to fatigue and reduced precision.

\textit{Ray+Hand} fell between the two, with moderate ratings in \textit{Physical Demand} and \textit{Frustration} (both medians = 4.0, SEM = 0.42) and better perceived performance than Ray alone.

Beyond the NASA-TLX scores, participants also provided open-ended responses on comfort and familiarity. When asked, “Which interaction felt more comfortable?” 53\% of participants selected \textit{HandOver}, citing its precision and ease of control. Another 35\% preferred \textit{Ray+Hand}, primarily praising its efficiency for mid-range tasks. Only 12\% favored \textit{Ray}, pointing to difficulties in accurately grabbing objects at the farthest distances.

For “Which interaction felt most familiar?”, 65\% again chose \textit{HandOver}, emphasizing that its mouse-like cursor was “easy to pick up quickly.” About 29\% preferred \textit{Ray+Hand} for its blend of “natural” hand gestures and partial cursor control. \textit{Ray} was selected by only 6\%, with participants describing it as “less intuitive” due to the need for continuous arm extension.
\vspace{-2mm}
\section{Applications}

HandOver enables a seamless transition between mouse-based input and embodied hand interaction, supporting a range of spatial computing tasks that benefit from both precision and expressivity. By decoupling control modalities across space and intent, HandOver extends existing workflows without introducing additional complexity or requiring users to abandon familiar input devices.
\begin{figure}[h]
  \centering
  \includegraphics[width=.8\columnwidth]{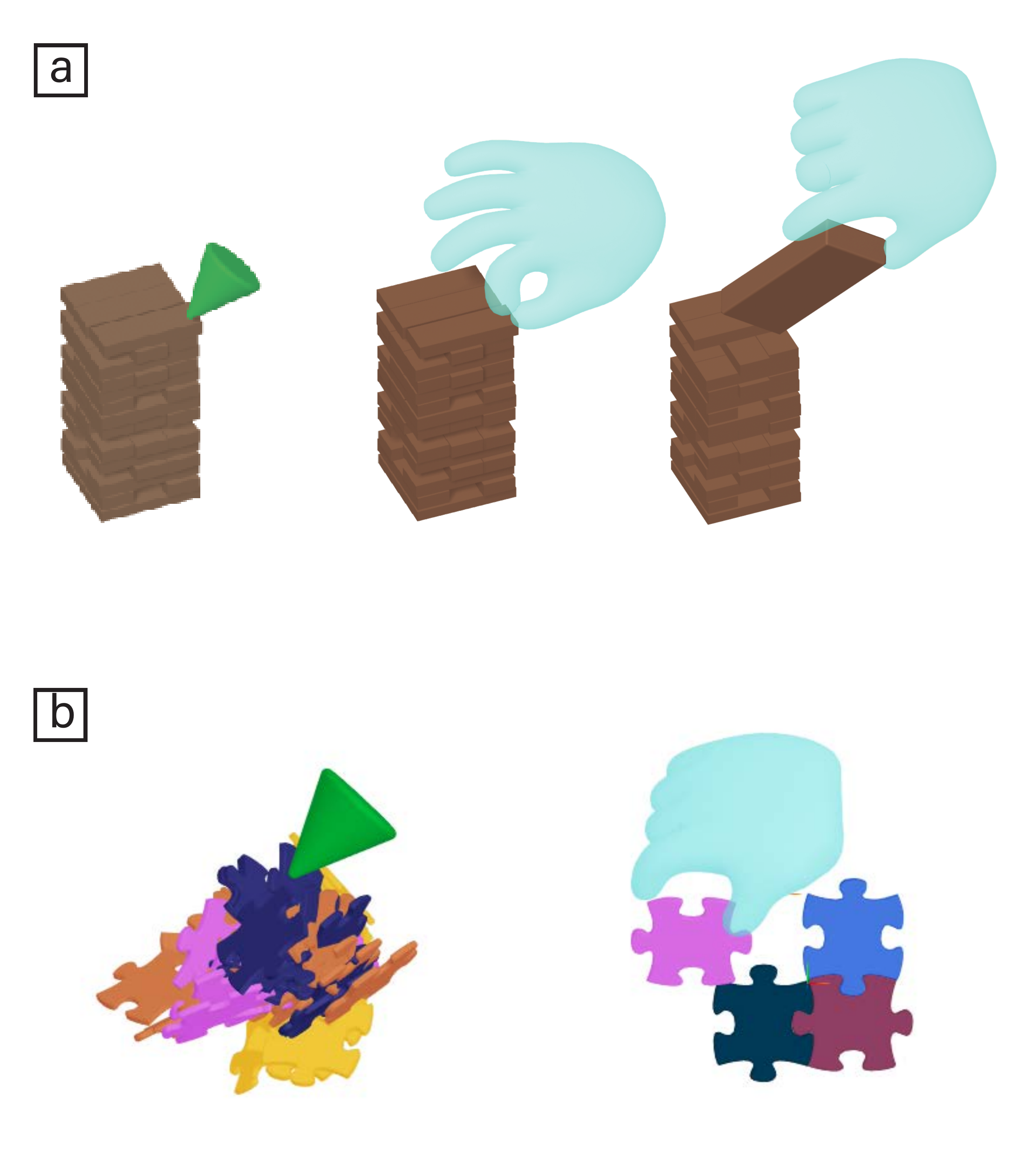}
  \caption{ Example application scenarios for HandOver across productivity and creative tasks. (a) Collaborative 3D modeling: users perform precise edits with the mouse and adjust object position using the hand clone.  (b) Virtual prototyping: users sketch or manipulate fine object details with the mouse while reorienting or scaling models in space with the hand clone, enabling fluid, bimanual coordination.}
  \label{fig:applications1}
\end{figure}
The following examples illustrate how HandOver supports diverse forms of spatial interaction, ranging from object manipulation to continuous control and bimanual workflows. Each subsection highlights a particular use case that benefits from the system’s ability to fluidly shift between precise cursor input and embodied hand interaction.

\subsection{Dexterous Pull and Docking}
Pulling or docking a physical prop with accuracy requires a high level of dexterity. However sometimes selecting the actual object in  a crowded VR scene can be cumbersome, specially when there is a lack of haptic feedback. HandOver affords that precision and dexterity to its users (Figure \ref{fig:applications1}).

\subsection{Continuous Trajectory Control}
\begin{figure}[h]
  \centering
  \includegraphics[width=.8\columnwidth]{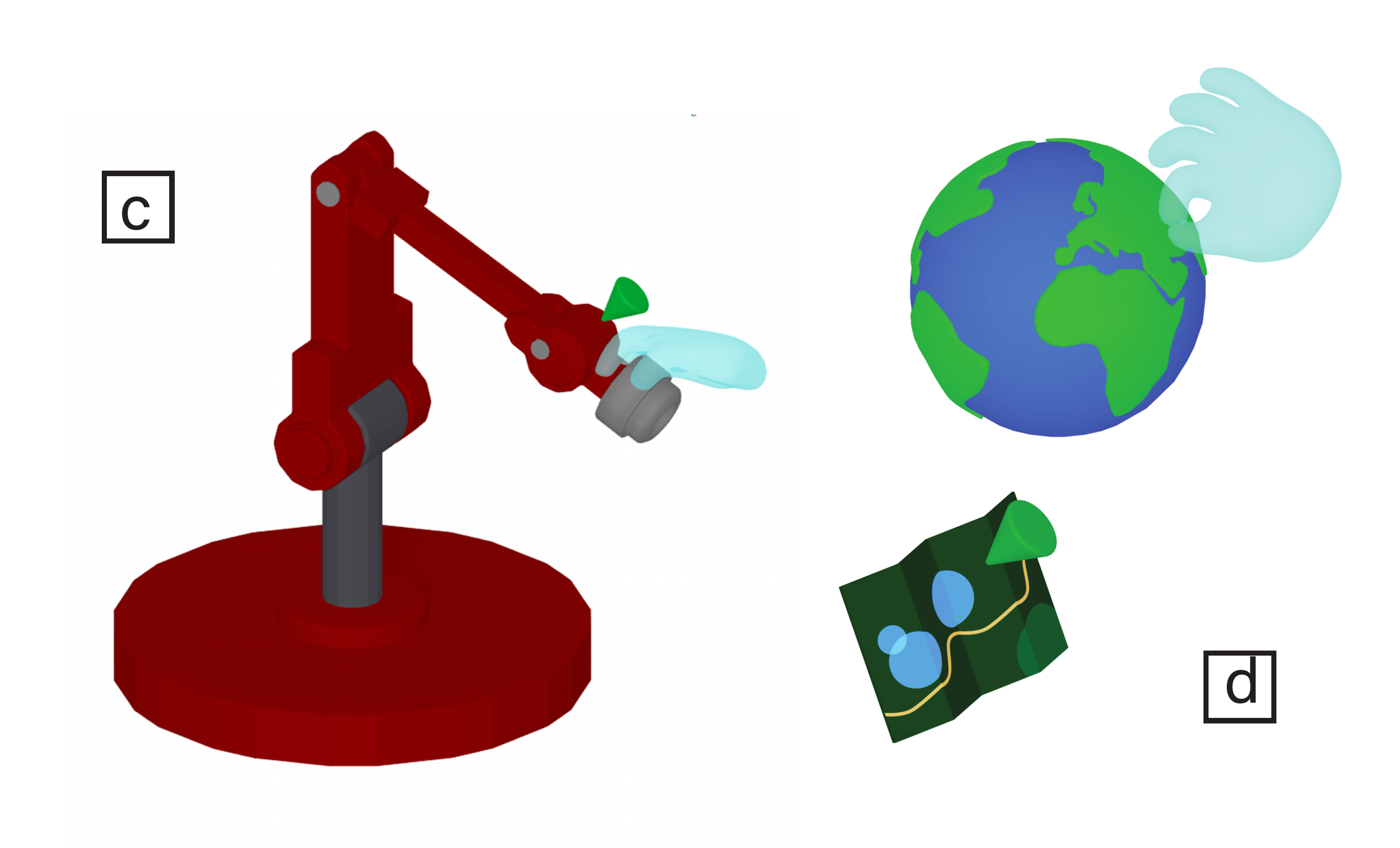}
  \caption{  (c) Remote assembly: distant components are selected with ray input and fine-tuned using mouse-based adjustments. (d) VR education: learners explore 2D-3D transtion models via mouse while demonstrating concepts with expressive hand gestures. }
  \label{fig:applications2}
\end{figure}

Often times a manipulation requires users to follow a precise trajectory, in situations like painting, robotic control or even gesture mapping for resizing (Figure \ref{fig:applications2}).

This can demonstrate critical also for collaborative 3D modeling environments, where users can select and edit geometry with high precision using the mouse, while using the hand clone to reposition or reorient objects. This allows for simultaneous micro-adjustments and macro-level spatial control.

\subsection{Bimanual interaction}
 In hybrid scenarios, users can employ mouse or ray selection to interact with distant components, then perform fine-grained alignment with a combination of the hand and the mouse input. 

\begin{figure}[h]
  \centering
  \includegraphics[width=.8\columnwidth]{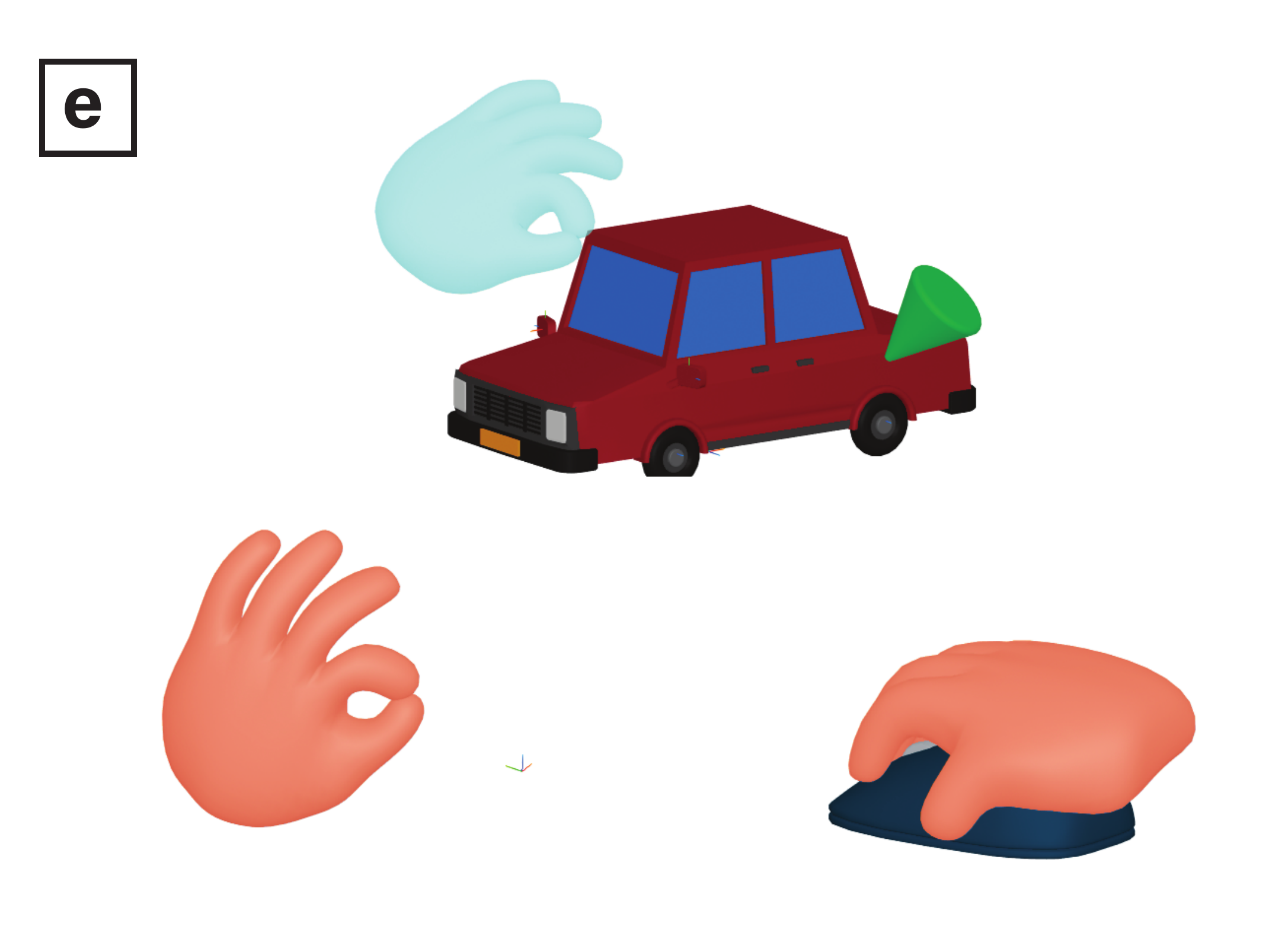}
  \caption{  (e) Bimanual spatial organization: the mouse is used to arrange or select fine-grained items while the hand clone groups or repositions container regions. }
  \label{fig:applications}
\end{figure}

HandOver further supports bimanual interaction. In spatial  \\workspace organization, users can use the mouse to manipulate individual files or interface elements while simultaneously using the non-dominant hand clone to cluster, pin, or resize spatial containers—mirroring bimanual workflows common in physical work environments . In creative and design contexts, such as virtual prototyping, one hand may control model orientation or scale while the other handles detailed sculpting or adjustment tasks, enabling fluid two-handed coordination without mode switching.

\section{Discussion}
HandOver demonstrated significant advantages in enabling intuitive distant object manipulation in XR. The depth-aware adaptive cursor and the hybrid interaction techniques allowed for precise targeting even at greater distances. Participants reported minimal cognitive load and enhanced task completion rates, particularly when using the HandOver for far-field tasks. Participants particularly benefited from the consistent control feel across distances, without needing to mentally switch strategies.

A key strength of the HandOver was its dynamic transition mechanism between the hand cursor and distant hand interaction modes. This allowed seamless switching between techniques without interrupting user flow, contributing to more natural interaction patterns.

Ray+Hand, combining Ray-based selection with HandOver-style manipulation, emerged as a compelling middle ground. Its performance approached that of HandOver during docking, especially at Near distances, while preserving the familiarity of ray-based pointing. However, the initial hand clone spawn point, centered at the cursor, may have slightly limited precision compared to the controlled placement in HandOver.

These findings suggest a promising design space: dynamically combining techniques like Ray+Hand and HandOver based on task demands could offer optimal performance across contexts. Qualitative feedback revealed that users appreciated the visual clarity provided by the distinct visual states of the hand clone and cursor. This reinforces the value of distinct interaction metaphors when switching modalities.

Future research could investigate multimodal feedback enhancements, such as audio cues or haptic vibrations, to further reinforce interaction states. Additionally, expanding the system to support bimanual interaction could open new possibilities for collaborative VR tasks.

\section{Limitations and Future Work}

While our results affirm the benefits of the HandOver technique, several limitations highlight opportunities for refinement. First, although the depth-aware interaction was generally effective, the built-in camera-based hand tracking occasionally led to inaccuracies, especially during fast movements or when hands left the optimal tracking zone. Future versions could integrate predictive tracking or use multi-camera setups to improve robustness.

We observed a reduction in precision when interacting with small targets at the farthest radial distance (6.0\,m). Adaptive gain scaling based on target size or distance may help mitigate this issue. While the circular layout in our task design enabled controlled comparisons, it does not fully capture the occlusion, clutter, and spatial irregularities present in real-world environments. Future work should explore performance under dynamic or crowded spatial scenarios. Participants also reported difficulty perceiving subtle changes in cursor opacity during rapid transitions. Enhancing visual contrast or introducing complementary modalities like haptic or audio cues could address this.

Although our interface supports full 3D depth input, object manipulation was restricted to a relatively small area surrounding the target, limiting the extent of remote repositioning. This reflects a bounded manipulation zone rather than true remote manipulation. In future work, we plan to evaluate performance in longer-range interaction tasks, such as reaching or placing across a room.

The study emphasized a seated workflow using a physical mouse, which enabled repeatable, fatigue-free precision input. Prior work suggests seated tasks remain common in XR productivity and design scenarios \cite{10.1145/3658407, grubert2020futurerevisitingmousekeyboard}. However, standing and mobile contexts warrant future validation.To extend beyond physical mice, we see potential in mid-air input alternatives such as TriPad \cite{10.1145/3613904.3642323} and XDTK \cite{10536201}, which emulate pointer-based control without requiring hardware. Integrating these methods could enable greater flexibility in user setups, while allowing dynamic switching between Ray+Hand and HandOver based on object size, distance, or ergonomics.

\section{Conclusion}
In this paper we have introduced HandOver, a hybrid interaction framework that seamlessly transitions from mouse‐based control to hand‐clone manipulation, enhancing both near‐field precision and far‐field naturalism in XR. By anchoring a depth‐aware 3D cursor to the user’s mouse input and automatically spawning a “virtual hand” when the user lifts off, HandOver substantially alleviates the fatigue and overshoot commonly observed in purely gestural or ray‐based approaches. Our comparative study against Ray+Hand and Ray confirms that integrating mouse control with intuitive hand interaction yields significantly higher precision, particularly in tasks that demand fine alignment or subtle movement across varying distances.

\balance
Beyond highlighting the ergonomic and performance benefits of hybrid solutions, our work underscores a broader message: no single interaction style suits every distance or object size. Instead, fluidly matching the interface to the user’s immediate demands, such as toggling from stable near‐field alignment to direct mid‐air grasp, better accommodates real‐world tasks. Future research can refine the bounding‐box detection logic, explore more adaptive gain factors, and study how bimanual or multimodal cues such as audio, haptics  might further improve user satisfaction and performance. Ultimately, HandOver reaffirms that bridging “the best of both worlds” stands as a promising direction for immersive 3D interfaces, ensuring smooth transitions in a single, cohesive workflow.

\bibliographystyle{ACM-Reference-Format}
\bibliography{sample-base}










\end{document}